\documentclass[fp,twocolumn]{jpsj3}
\usepackage{txfonts}
\usepackage{amsmath,amssymb}
\usepackage{graphicx}
\usepackage{color}

\title{Effect of Parallel Magnetic Field on Superconductivity of Ultrathin Metal Films
Grown on a Cleaved GaAs Surface}

\author{Takayuki Sekihara, Takahiro Miyake, Ryuichi Masutomi, and Tohru Okamoto}
\inst{Department of Physics, University of Tokyo, 7-3-1 Hongo, Bunkyo-ku, Tokyo 113-0033, Japan}

\abst{
The parallel-magnetic-field $H_\parallel$ dependence of the superconducting transition temperature $T_c$ is studied
for ultrathin films of In, Bi, and Al grown on GaAs(110).
In the case of In films in the monolayer regime,
$T_c$ exhibits a quadratic-like $H_\parallel$ dependence,
which is one order of magnitude stronger than that previously observed in monolayer Pb films
by the present authors [Phys. Rev. Lett. {\bf 111}, 057005 (2013)].
The results are well reproduced by a model
developed for an inhomogeneous two-dimensional superconducting state
in the presence of a moderate Rashba spin-orbit interaction.
The Rashba spin splitting is estimated to be 0.04~eV,
which is much smaller than that expected for monolayer Pb films.
In a few-monolayer Bi film,
the suppression of $T_c$ with increasing $H_\parallel$ is comparable to that in monolayer Pb films.
On the other hand, much stronger suppression, which is attributed to the Pauli paramagnetic effect,
was observed for the Al film.
}

\begin{document}
\maketitle

\section{Introduction}

Recently it has been demonstrated that 
superconductivity can occur even in one-atomic-layer metal films.
\cite{Zhang2010,Uchihashi2011}
In an ultrathin film grown on a substrate,
the asymmetry of the confining potential is expected to cause
the Rashba spin-orbit interaction.\cite{Rashba1960,Bychkov1984}
It was shown in Ref.~\citen{Barzykin2002} that
the superconducting critical magnetic field in the parallel direction,
where the orbital effect is absent,
can be very high for a two-dimensional system with a large Rashba spin splitting $\Delta_R$
because of the formation of an inhomogeneous superconducting state 
similar to the Fulde-Ferrell-Larkin-Ovchinnikov state.\cite{Fulde1964,Larkin1965}
This Fulde-Ferrell-Larkin-Ovchinnikov-like state is easily destroyed by disorder.\cite{Dimitrova2003,Dimitrova2007}
Instead, a different inhomogeneous superconducting state appears, which is called the long-wavelength helical state.
\cite{Dimitrova2003,Dimitrova2007}
Here the Cooper pairs have a very small but nonzero momentum $\bold{q}$
and the order parameter varies as $\exp (i \bold{q} \cdot \bold{r})$.
Very recently, the present authors studied the effect of a parallel magnetic field $H_\parallel$
on the superconductivity of monolayer Pb films on GaAs(110).\cite{Sekihara2013}
Superconductivity was found to occur even for $H_\parallel=14$~T,
which is much higher than the Pauli paramagnetic limiting field $H_P$.\cite{Clogston1962,Chandraskhar1962}
The observed weak $H_\parallel$ dependence of the superconducting transition temperature $T_c$ is well reproduced by 
\begin{equation}
T_c=T_{c0}- \frac{\pi \tau}{2 k_B \hbar} \left( \mu _B H_\parallel \right)^2,
\end{equation}
where $T_{c0}$ is the zero-field value of $T_c$,
$\tau$ is the elastic scattering time, and $\mu_B$ is the Bohr magneton.
This expression is derived from the theory of Refs.~\citen{Dimitrova2003} and \citen{Dimitrova2007}
developed for the case $\Delta_R \gg \hbar \tau^{-1}$.

The strength of the spin-orbit interaction strongly depends on the atomic number $Z$.
In order to confirm the origin of the robustness of superconductivity against $H_\parallel$
in Pb films ($Z=83$),
it is important to study ultrathin superconducting films of other materials.
In this work, we extend our studies to ultrathin films of In
($Z=49$), Bi ($Z=83$), and Al ($Z=13$).
In the case of In films, superconductivity was observed in the monolayer regime.
The $H_\parallel$ dependence of $T_c$ is one order of magnitude stronger than that in the Pb films
and does {\it not} quantitatively agree with Eq.~(1).
Since $\Delta_R$ is expected to be small in the In films,
we also extend the analysis of Refs.~\citen{Dimitrova2003} and \citen{Dimitrova2007}
to the case where $\Delta_R$ is comparable to or smaller than $\hbar \tau^{-1}$.
The experimental results are well reproduced by the calculation
with $\Delta_R \approx 0.04$~eV, which is much smaller than that expected for the Pb films.
For Bi and Al, superconducting monolayer films were not obtained.
We measured the $H_\parallel$ dependence of $T_c$ for 
the smallest thickness for which superconductivity was observed.
In the Bi film, it is as weak as that observed in the monolayer Pb films.
In the Al film, on the other hand, it is strong and
can be explained in terms of the Pauli paramagnetic effect.

\section{Experimental Methods}

The sample preparation methods and experimental setup are similar to those previously described.\cite{Sekihara2013}
We used a nondoped insulating GaAs single-crystal substrate
so as not to create conduction channels in the substrate.
Current and voltage electrodes were prepared at room temperature
by the deposition of gold films onto noncleaved surfaces.
The cleavage of GaAs, the deposition of an ultrathin film by quench condensation, and resistance measurements 
were performed at low temperatures in an ultrahigh-vacuum chamber immersed in liquid He.
The amount deposited was measured with a quartz crystal microbalance
and determined with an accuracy of about 5\%.
The four-probe resistance of the ultrathin film on a cleaved GaAs(110) surface (4 $\times$ 0.35 mm${}^2$)
was measured using the standard lock-in technique at 13.1 Hz.
The magnetic-field direction with respect to the surface normal
was precisely controlled using a rotatory stage on which the sample was mounted,
together with a Hall generator, a RuO$_2$ resistance thermometer, and a heater.
The sample stage can be cooled to 0.5~K via a silver foil linked to a pumped ${}^3$He refrigerator.
All the data were taken when the temperature of the sample stage was kept constant
so as to ensure thermal equilibrium between the sample and the thermometer.
The magnetoresistance effect of the RuO$_2$ resistance thermometer was systematically calibrated
against the vapor pressure of the liquid ${}^3$He or ${}^4$He for various temperatures.
After the correction, $T_c$ can be determined with a relative accuracy of better than 0.2\%.\cite{RuO2}

\section{Results and Discussion}

\subsection{Thickness dependence}

Figure 1 shows the zero-field superconducting transition temperature $T_{c0}$
and the normal-state sheet resistance $R_N$
as a function of the nominal thickness $d$ of quench-condensed films of In, Al, Pb, and Bi.
We defined the transition temperature as the temperature at which the sheet resistance $R_{sq}$ reaches $0.5 R_N$.
Compared with the results on a glazed alumina substrate coated with amorphous Ge reported in Ref.~\citen{Haviland1989},
$R_N$ is lower and $T_{c0}$ is higher for Pb and Bi.\cite{Sekihara2013}
We attribute this to the atomically flat surface of the cleaved GaAs substrate.
Films of In and Pb show superconductivity even in the monolayer regime ($d \lesssim 0.3$~nm).
For Al and Bi, on the other hand, we could not attain a superconducting monolayer film.
The morphology of the Al film will be discussed in Sect.~3.4.

\begin{figure}[t]
\begin{center}
\includegraphics[width=0.85\linewidth, clip]{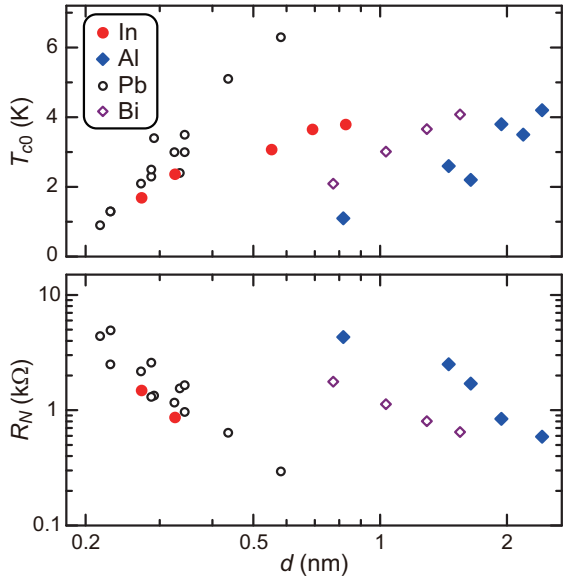}
\caption{(Color online) 
Zero-field superconducting transition temperature
and normal-state sheet resistance as a function of
the nominal thickness of quench-condensed films on GaAs(110).
Data for Pb and Bi are from Ref.~\citen{Sekihara2013}.
}
\end{center}
\end{figure}

\subsection{In films}

Figure 2 shows the temperature dependence of the sheet resistance $R_{sq}$ 
of In films with atomic areal densities $n=10.4$ and 12.5~nm${}^{-2}$.
The corresponding values of $d$ are 0.27 and 0.32~nm, respectively,
with a bulk density of 38.3~nm${}^{-3}$.
Since they are close to the cube root of the volume per atom in bulk indium (0.30~nm),
the films are considered to be almost monolayer.
In contrast to epitaxially grown films,\cite{Zhang2010,Uchihashi2011}
the quench-condensed films are expected to be amorphous or highly disordered.\cite{Haviland1989}
The normal-state sheet resistances $R_N$ in our In films
are higher than 410~$\Omega$, the value for the monolayer In film grown epitaxially on Si(111)
by Uchihashi {\it et al.}\cite{Uchihashi2011}
As in the case of the Pb films,\cite{Sekihara2013}
superconductivity is observed even for $H_\parallel$
higher than the Pauli paramagnetic limiting field
$H_P ({\mathrm T}) =1.86 T_{c0} ({\mathrm K})$.\cite{Clogston1962,Chandraskhar1962}
However, the effect of $H_\parallel$ on $T_c$ is much stronger than that in the Pb films.

\begin{figure}[t]
\begin{center}
\includegraphics[width=0.85\linewidth, clip]{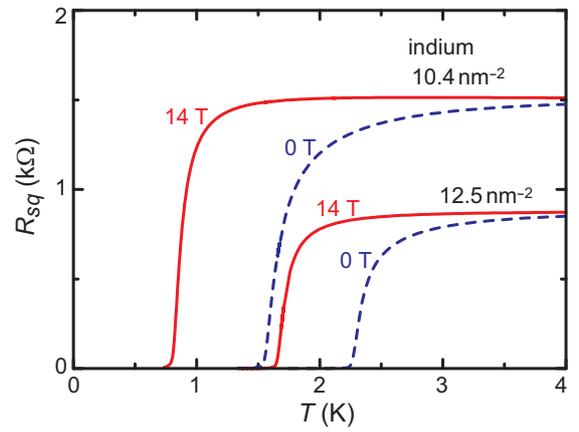}
\caption{(Color online)
$T$ dependence of the sheet resistance of In films for two atomic areal densities.
Dashed (blue) curves are obtained at zero magnetic field.
Solid (red) curves are obtained in a parallel magnetic field of 14~T.
}
\end{center}
\end{figure}

In the presence of the perpendicular component $H_\perp$ of the magnetic field,
the superconducting state is easily destroyed because of the orbital effect.
Figure 3 shows the $H_\perp$ dependence of $R_{sq}$ for
the In film with $n=12.5~\mathrm{nm}^{-2}$ at $T=0.83$~K.
Data obtained for $H_\parallel=0$ (solid curve) and $H_\parallel \approx H=14$~T (filled circles) are compared.
While the $H_\perp$ dependence of $R_{sq}$ was found to be almost independent of the presence of $H_\parallel$
in the case of the Pb films,\cite{Sekihara2013}
the suppression of superconductivity by $H_\parallel$ is apparent for the In films.
Since $H_\parallel$ does not change the Lorentz force acting on vortices,
the difference in $R_{sq}$ for low $H_\perp$ is attributed to
that in the pinning force, which should be related to the superconducting gap.

\begin{figure}[bt]
\begin{center}
\includegraphics[width=0.85\linewidth, clip]{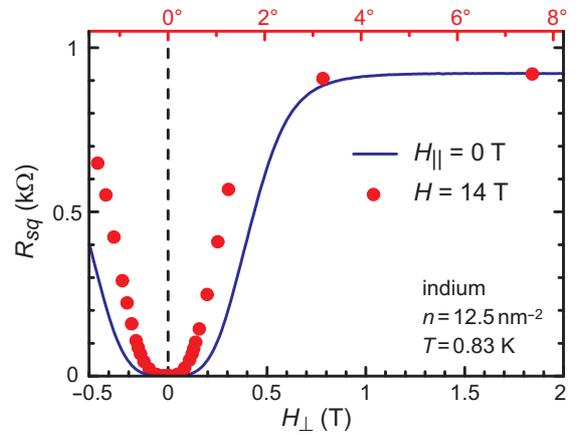}
\caption{(Color online)
$H_\perp$ dependence of the sheet resistance of the In film with $n=12.5~\mathrm{nm}^{-2}$ at $T=0.83$~K.
The solid (blue) curve is obtained for the perpendicular-magnetic-field direction ($H_\parallel=0$).
Filled (red) circles are obtained by changing the magnetic-field angle (upper axis)
at a fixed strength of 14~T.
}
\end{center}
\end{figure}

In Fig.~4, the $H_\parallel$ dependence of $\Delta T_c \equiv T_c-T_{c0}$ is shown for the In films.
Data for the Pb films\cite{Sekihara2013} are also plotted for comparison.
As in the Pb films, $T_c$ exhibits a quadratic-like $H_\parallel$ dependence,
which is at least qualitatively consistent with Eq.~(1).
However, the suppression of $T_c$ is one order of magnitude stronger than that in the Pb films.
If Eq.~(1) is applied to the experimental data,
$\tau$ is determined to be 43 and 35~fs
for $n=10.4$ and $12.5~\mathrm{nm}^{-2}$, respectively.
These values are significantly larger than
$\tau=2.8$ and 4.0~fs, roughly estimated from $R_N$.\cite{ScatteringTime}
We attribute this discrepancy to the fact that
Eq.~(1) is derived for $\Delta_R \gg \hbar \tau^{-1}$.
Since the Rashba spin splitting originates both from the asymmetry of the confining potential
and atomic spin-orbit coupling,
$\Delta_R$ is expected to be small for elements with small atomic numbers $Z$.
In the following, we consider the case where $\Delta_R$ is smaller than $\hbar \tau^{-1}$.

\begin{figure}[t]
\begin{center}
\includegraphics[width=0.85\linewidth, clip]{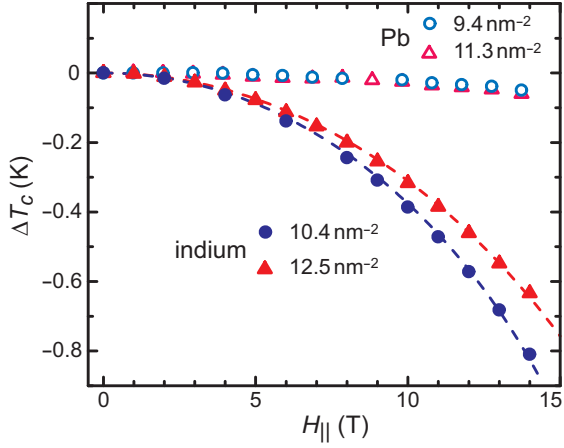}
\caption{(Color online) 
$H_\parallel$ dependence of $\Delta T_c \equiv T_c-T_{c0}$.
Solid symbols represent data for the In films.
Open symbols are data from Ref.~\citen{Sekihara2013} for the Pb films.
The dashed lines represent the best fits to Eq.~(6).
}
\end{center}
\end{figure}

The stability of the long-wavelength helical state\cite{Dimitrova2003,Dimitrova2007}
can be interpreted as being due to mixing of the spin-up state and spin-down state,
which weakens the pair-breaking effect of $H_\parallel$.\cite{nonzeroq}
For a qualitative understanding, it is helpful to use the analogy with the theory
developed for the case where the spin mixing is caused by
the spin-orbit scattering of impurities.\cite{Maki1966,Klemm1975}
For a superconductor with strong spin-orbit scattering,
$T_c$ is given by
\begin{equation}
\ln \left( \frac{T_c}{T_{c0}} \right)+\psi \left( \frac{1}{2} +
\frac{3 \tau_\mathrm{SO} (\mu_B H_\parallel)^2 }{4 \pi \hbar k_B T_c}
\right) - \psi \left( \frac{1}{2}\right) =0,
\end{equation}
where $\tau_\mathrm{SO}$ is the spin-orbit scattering time and $\psi(x)$ is the digamma function.
For $T_c \approx T_{c0}$, 
the $H_\parallel$ dependence of $T_c$ is obtained as
\begin{equation}
T_c=T_{c0}- \frac{3 \pi \tau_\mathrm{SO}}{8 k_B \hbar} \left( \mu _B H_\parallel \right)^2.
\end{equation}
In the present case, we consider nonmagnetic elastic scattering by defects,
which conserves the spin of the conduction electrons.\cite{Sekihara2013}
However, the subsequent spin precession due to the Rashba effect 
can change the spin direction.
If the precession frequency $\Omega_R =\hbar^{-1} \Delta_R$ 
is high enough ($\Omega_R \tau \gg 1$),
the effective spin relaxation time $\tau_{s}^\ast$
is expected to be comparable to $\tau$.
Actually, Eq.~(3) is similar to Eq.~(1) 
when $\tau_\mathrm{SO}$ is replaced by $\tau_{s}^\ast \sim \tau$,
while the origin of the spin relaxation is different.
For $\Omega_R \tau \ll 1$, on the other hand,
the electron spin rotates by a small angle $\Omega_R \tau$ between successive scattering events
and follows a random walk.
This is the D'yakonov-Perel mechanism\cite{Dyakonov1971,Zutic2004}
and $\tau_{s}^\ast \sim (\Omega_R{}^2 \tau)^{-1} \gg \tau$ is obtained.
In this case, by analogy with Eq.~(3), the coefficient of the quadratic $H_\parallel$ dependence of $T_c$
can be much larger than that in Eq.~(1).

For a quantitative description, we here extend the analysis of Refs.~\citen{Dimitrova2003}
and \citen{Dimitrova2007}
to the case $k_B T_{c0} \ll \Delta_R \lesssim \hbar \tau^{-1} \ll \epsilon_F$,
where $\epsilon_F$ is the Fermi energy.
We take into account Cooper blocks\cite{Dimitrova2007} where the chiralities of the two electrons are different,
while their contributions are negligible for $\Delta_R \gg \hbar \tau^{-1}$.
Equation (13) of Ref.~\citen{Dimitrova2003},
which represents the Cooper kernel\cite{Dimitrova2003,Dimitrova2007} as a function of the Matsubara frequency $\omega$
and is equivalent to Eq.~(94) of Ref.~\citen{Dimitrova2007},
is replaced by
\begin{equation}
K \left( \omega \right)=4 \tau \frac{
I_{s}^{0}
\left[1-I_{s}^{2}-J \right]
+\left(I_{a}^{1} \right)^2}{
\left(1-I_{s}^{0} \right)
\left[1-I_{s}^{2}-J \right]
-\left(I_{a}^{1} \right)^2},
\end{equation}
where we introduce
\begin{equation}
J=\frac{\hbar^2}{4\tau} \frac{\bar{\omega}}{\hbar^2 \bar{\omega}^2 +(\alpha p_F)^2}
=\frac{1}{\tau} \frac{\bar{\omega}}{4 \bar{\omega}^2 +\Omega_R{}^2},
\end{equation}
with $\bar{\omega}=\omega+1/2\tau$.
After some algebra, we obtain
\begin{equation}
\frac{T}{T_{c0}}=\exp \left[ -\frac{2\pi k_B T}{\hbar} \sum_{\omega>0}^{\infty} \frac{1}{\omega}
\frac{2\tau_{s}^\ast h_\parallel{}^2}{\omega^2 \tau'+\omega+2\tau_{s}^\ast h_\parallel{}^2}
\right],
\end{equation}
with $h_\parallel=\mu_B H_\parallel/\hbar$, $\tau'=4/\Omega_R{}^2 \tau$, and
\begin{equation}
\tau_{s}^\ast=
\left( 1+\frac{1}{2 \Omega_R{}^2 \tau^2} \right) \tau.
\end{equation}
By solving Eq.~(6) for $T$, $T_c$ is numerically determined as a function of $H_\parallel$.
For small magnetic fields ($|\Delta T_c| \ll T_{c0}$), it is given by
\begin{equation}
T_c=T_{c0}- \frac{\pi \tau_{s}^\ast f(a)}{2 k_B \hbar} \left( \mu _B H_\parallel \right)^2,
\end{equation}
where
\begin{equation}
f(a)=
1-\frac{2a}{\pi^2} \left[\psi \left(\frac{1}{2}+\frac{1}{a} \right)
-\psi \left( \frac{1}{2} \right) \right]
\end{equation}
with
\begin{equation}
a=\frac{8\pi \hbar k_B T_{c0}}{\tau \Delta_R{}^2}.
\end{equation}
For $\Delta_R \gg \hbar \tau^{-1}$ ($\Omega_R \tau \gg 1$),
we have $\tau_{s}^\ast \approx \tau$, $a \approx 0$, $f(a) \approx 1$,
and then Eq.~(8) agrees with Eq.~(1).
This condition is expected to be satisfied for the Pb films
since the values of $\tau$ estimated from $R_N$
are in good agreement with those obtained from
the $H_\parallel$ dependence of $T_c$ using Eq.~(1).
\cite{Sekihara2013}
In the opposite limit $\Delta_R \ll \hbar \tau^{-1}$,
we obtain $\tau_{s}^\ast \approx (2 \Omega_R{}^2 \tau)^{-1}$,
which is consistent with the D'yakonov-Perel model.
For $\Delta_R \ll \hbar \tau^{-1}$ and
$\Delta_R{}^2 \lesssim (k_B T_{c0})(\hbar \tau^{-1})$,
the reduction of $f(a)$, which arises from the $\omega^2 \tau'$ term in Eq.~(6),
should be taken into account.

The dashed lines in Fig.~4 are the best fits to Eq.~(6).
Here we use $\tau=2.8$ and 4.0~fs, which are estimated from $R_N$.
Our calculation well reproduces the experimental results.
The obtained values of $\Delta_R$ are 39 and 35~meV
for the In films with $n=10.4$ and $12.5~\mathrm{nm}^{-2}$, respectively.
\cite{uncertainty}
They are one order of magnitude smaller than those measured for
monolayer Bi (Refs.~\citen{Gierz2009,Hatta2009,Sakamoto2009})
and Pb (Ref.~\citen{Yaji2010}) films grown on Si(111) or Ge(111)
by angle-resolved photoelectron spectroscopy.
For our Pb films on GaAs(110),\cite{Sekihara2013}
$\Delta_R$ is expected to be larger or at least comparable to $\hbar \tau^{-1} \sim 0.2$~eV
since the observed $H_\parallel$ dependence of $T_c$ agrees with Eq.~(1), which is derived for $\Delta_R \gg \hbar \tau^{-1}$.
It has been shown by a tight-binding calculation that
the Rashba parameter is proportional to the product of
the gradient of the surface potential and the magnitude of the atomic spin-orbit coupling.
\cite{Petersen2000}
Since the latter scales as $Z^4$,
it seems likely that $\Delta_R$ varies by one order of magnitude from In ($Z=49$) to Pb ($Z=82$) or Bi ($Z=83$).

\subsection{Bi films}

Figure 5 shows the $H_\parallel$ dependence of $T_c$
for the Bi film with $d=0.77$~nm,
the smallest thickness for which superconductivity was observed.
Since this value is a few times larger than 
the cube root of the volume per atom in bulk Bi (0.33~nm),
the film is not considered to be a monolayer.
Thus, it is not appropriate to apply the theory developed for two-dimensional systems.
On the other hand, the $H_\parallel$ dependence of $T_c$
was found to be very weak.
The suppression of $T_c$ is larger, but comparable to
that in the Pb films.~\cite{Sekihara2013}
The Rashba spin-orbit interaction may also play an essential role in the ultrathin
film of Bi with $Z=83$, which is close to $Z=82$ for Pb.

\begin{figure}[t]
\begin{center}
\includegraphics[width=0.85\linewidth, clip]{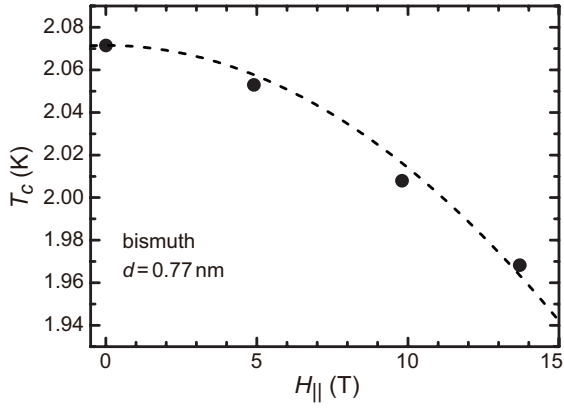}
\caption{
$H_\parallel$ dependence of $T_c$ for the Bi film with $d=0.77$~nm.
The dashed line is the best quadratic fit.
}
\end{center}
\end{figure}

\subsection{Al films}

In the case of Al films, the smallest thickness for the observation of superconductivity was 0.82~nm,
which is larger than the cube root of the volume per atom in bulk Al (0.26~nm).
At $d=0.68$~nm, $R_{sq}$ exhibits an insulating temperature dependence
as shown in the inset of Fig.~6.
In order to discuss the morphology of the quench-condensed Al films,
we study the magnetoresistance of the insulating film in the {\it perpendicular} direction.
In Fig.~6, the data at $T=0.58$~K are shown together with those for the Pb film
in the submonolayer regime ($d=0.16$~nm).
While the $H_\perp$ dependence of $R_{sq}$ is positive for the Pb film,
it is negative for the Al film.
Such strong negative magnetoresistance has been reported for granular Pb films
on fire-polished glass or quartz substrates,
where the electrical conduction is dominated
by quasiparticle tunneling between superconducting grains.\cite{Barber2006}
The negative magnetoresistance in the insulating regime is explained as a result of the enhancement of the conductance
due to the suppression of the superconducting energy gap by the magnetic field.
It seems plausible to suppose that the negative magnetoresistance observed in the present Al film
results from the same mechanism and that the morphology is granular,\cite{Liu1993}
while the monolayer Pb films are considered to be homogeneously disordered on GaAs(110).
On the other hand, the negative magnetoresistance in the Al film survives even in high magnetic fields
where the superconducting state is expected to be destroyed.
Further study is required to understand the origin of the negative magnetoresistance
in the insulating regime.

\begin{figure}[t]
\begin{center}
\includegraphics[width=0.85\linewidth, clip]{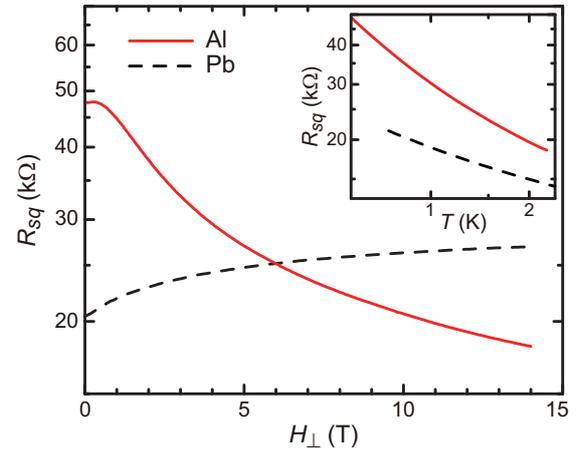}
\caption{(Color online) 
$H_\perp$ dependence of the sheet resistance.
The solid (red) curve is obtained for the Al film with $d=0.68$~nm at $T=0.58$~K.
The dashed curve is obtained for the Pb film with $d=0.16$~nm ($n=5.4~\mathrm{nm}^{-2}$) at $T=0.83$~K.
The inset shows the $T$ dependence of $R_{sq}$ at $H=0$.
}
\end{center}
\end{figure}

Figure 7 shows the $H_\parallel$ dependence of $T_c$
for the Al film with $d=0.82$~nm.
In contrast to the case of the Bi film shown in Fig.~5,
$T_c$ decreases rapidly with increasing $H_\parallel$.
Since the spin-orbit interaction is very small in Al ($Z=13$),
the critical magnetic field is expected to be determined by the Pauli paramagnetic limit.
The phase transition due to the paramagnetic effect is of the second order
for $T_c \geq 0.56 T_{c0}$.\cite{Sarma1963,Maki1964} 
The calculated phase boundary shown as the dashed curve in Fig.~7
qualitatively reproduces the experimental behavior.
While the energy gap at $T=0$ is given by
$\Delta_0=1.76 k_B T_{c0}$ in the theory,
it was shown in Ref.~\citen{Sacepe2008} that
disorder can lead to an increase in the $\Delta_0/k_B T_{c0}$ ratio.
If this is the case, the phase boundary may shift toward a higher $H_\parallel$.
The dotted curve was obtained by multiplying $H_\parallel$ for the dashed curve by 1.44
so as to fit the experimental data.

\begin{figure}[t]
\begin{center}
\includegraphics[width=0.85\linewidth, clip]{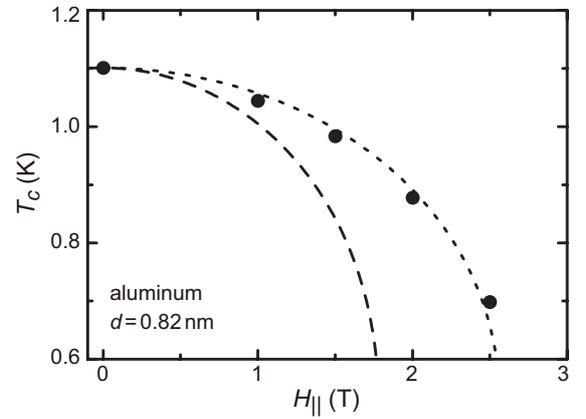}
\caption{
$H_\parallel$ dependence of $T_c$ for the Al film with $d=0.82$~nm.
The dashed curve is the second-order phase transition lines
calculated from Ref.~\citen{Maki1964}.
The dotted curve is obtained from the dashed curve
by multiplying its abscissas by 1.44.
}
\end{center}
\end{figure}

\section{Summary}

As reported in this paper, we have studied the effect of a parallel magnetic field
on the superconductivity of ultrathin metal films
grown on a cleaved GaAs(110) surface.
In monolayer In films, the $H_\parallel$ dependence of $T_c$ is much stronger than
that expected from Eq.~(1), which was derived for $\Delta_R \gg \hbar \tau^{-1}$.
We extended the analysis of Refs.~\citen{Dimitrova2003} and \citen{Dimitrova2007}
to the case $\Delta_R \lesssim \hbar \tau^{-1}$.
The theory well reproduces the experimental results when we use $\Delta_R \approx 0.04$~eV,
which is one order of magnitude smaller than that expected for monolayer Pb films.
In a few-monolayer Bi film, the suppression of $T_c$ with increasing $H_\parallel$ is very weak and 
comparable to that in the monolayer Pb films.
In the Al film, on the other hand, $T_c$ decreases rapidly with increasing $H_\parallel$
as expected from the simple calculation of the Pauli paramagnetic effect.
By comparing the experimental results for different $Z$,
it has been confirmed that a strong spin-orbit interaction is essential for 
the robustness of superconductivity against $H_\parallel$.

\begin{acknowledgment}

This work has been partially supported by Grants-in-Aid for Scientific Research
(A) (No. 21244047) and (B) (No. 26287072) from MEXT, Japan.

\end{acknowledgment}


\begin{thebibliography}{99}
\bibitem{RuO2}
The RuO$_2$ resistance thermometer (RO-600) was manufactured by Scientific Instruments, Inc.,
which offers a simple linear formula for a positive magnetoresistance.
It allows the correction of any apparent temperature,
measured in a magnetic field of up to 16~T, to the actual temperature
with an error of less than $\pm 1.6$\%
for a broad temperature range of 0.036 to 4.2~K.
Our calibration agrees with the formula within the error bars.
\bibitem{uncertainty}
Note that these values have some uncertainty due to
the uncertainty in the estimation of $\tau$.
\end{thebibliography}
\end{document}